\documentclass[conference]{IEEEtran}
\IEEEoverridecommandlockouts
\usepackage{amsmath,amssymb,amsfonts}
\usepackage{algorithmic}
\usepackage{todonotes}
\usepackage{graphicx}
\usepackage{textcomp}
\usepackage{colortbl}
\usepackage{url}
\usepackage{multirow}
\usepackage{hhline}
\usepackage{float}
\usepackage{authblk}
\usepackage{dblfloatfix}

\usepackage[backend=biber,style=numeric,natbib=true,sorting=none,url=false]{biblatex}
\usepackage[none]{hyphenat}
\def\BibTeX{{\rm B\kern-.05em{\sc i\kern-.025em b}\kern-.08em
    T\kern-.1667em\lower.7ex\hbox{E}\kern-.125emX}}

\renewcommand*{\figurename}{Fig.} %change Fig. to Abb.
\bibliography{refs.bib}
\begin{document}

\title{Open or not open: Are conventional radio access networks more secure and trustworthy than Open RAN? 
{\footnotesize \textsuperscript{}}
%\thanks{The authors acknowledge the financial support by the Federal Ministry of Education and Research of Germany in the program of "Souverän. Digital. Vernetzt." Joint project 6G-RIC, project identification number: 16KISK020K and 16KISK0[21-35].}
}

\author[1]{Felix Klement}
\author[1]{Stefan Katzenbeisser}
\author[2]{Vincent Ulitzsch}
%\author[2]{Jean-Pierre Seifert}
\author[3]{Juliane Krämer}
\author[4]{\authorcr Slawomir Stanczak}
\author[4]{Zoran Utkovski}
\author[4]{Igor Bjelakovic}
\author[5]{Gerhard Wunder}

\affil[1]{Computer Engineering, University of Passau, 94032 Passau, Germany}
\affil[2]{Security in Telecommunications, TU Berlin, 10587 Berlin, Germany}
\affil[3]{Data Security and Cryptography, University of Regensburg, 93053 Regensburg, Germany}
\affil[4]{Fraunhofer Heinrich Hertz Institute, 10587 Berlin, Germany}
\affil[5]{Cybersecurity and AI Group, FU Berlin, 14195 Berlin, Germany}

\maketitle

\begin{abstract}
%The Open-RAN architecture is a highly promising and future-oriented architecture. It is intended to open up the radio access network and enable more innovation and competition in the market. This will lead to RANs for current 5G networks, but especially for future 6G networks, to move away from the current centralised, provider-specific 3G RAN architecture and therefore even better meet the requirements for future RANs. However, the change in design has also created a drastic shift in the attack surface compared to conventional RANs. In the past, this has often led to negative headlines, which in summary have often associated O-RAN with faulty or inadequate security. In this paper, we analyze what components are involved in an Open-RAN deployment, how the current state of security is to be assessed and what measures need to be taken to ensure secure operation.

The Open RAN architecture is a promising and future-oriented architecture. It is intended to open up the radio access network (RAN) and enable more innovation and competition in the market. This will lead to RANs for current 5G networks, but especially for future 6G networks, evolving from the current highly integrated, vendor-specific RAN architecture towards disaggregated architectures with open interfaces that will enable to better tailor RAN solutions to the requirements of 5G and 6G applications. However, the introduction of such an open architecture substantially broadens the attack possibilities when compared to conventional RANs. In the past, this has often led to negative headlines that in summary have associated Open RAN with faulty or inadequate security. In this paper, we analyze what components are involved in an Open RAN deployment, how to assess the current state of security, and what measures need to be taken to ensure secure operation.
\end{abstract}


\section{Introduction to Open-RAN}
\subsection{Motivation}
\label{motivation}

Modern communication is one of the central pillars of successful digitization. Particularly instrumental is the recently introduced 5G technology and its ongoing evolution towards 6G. In addition to public mobile networks, 5G technology - and in the long term 6G technology - will also be used for local radio networks (so-called private networks or campus networks). 

\begin{figure}
	\centering
	\includegraphics[width=0.5\textwidth]{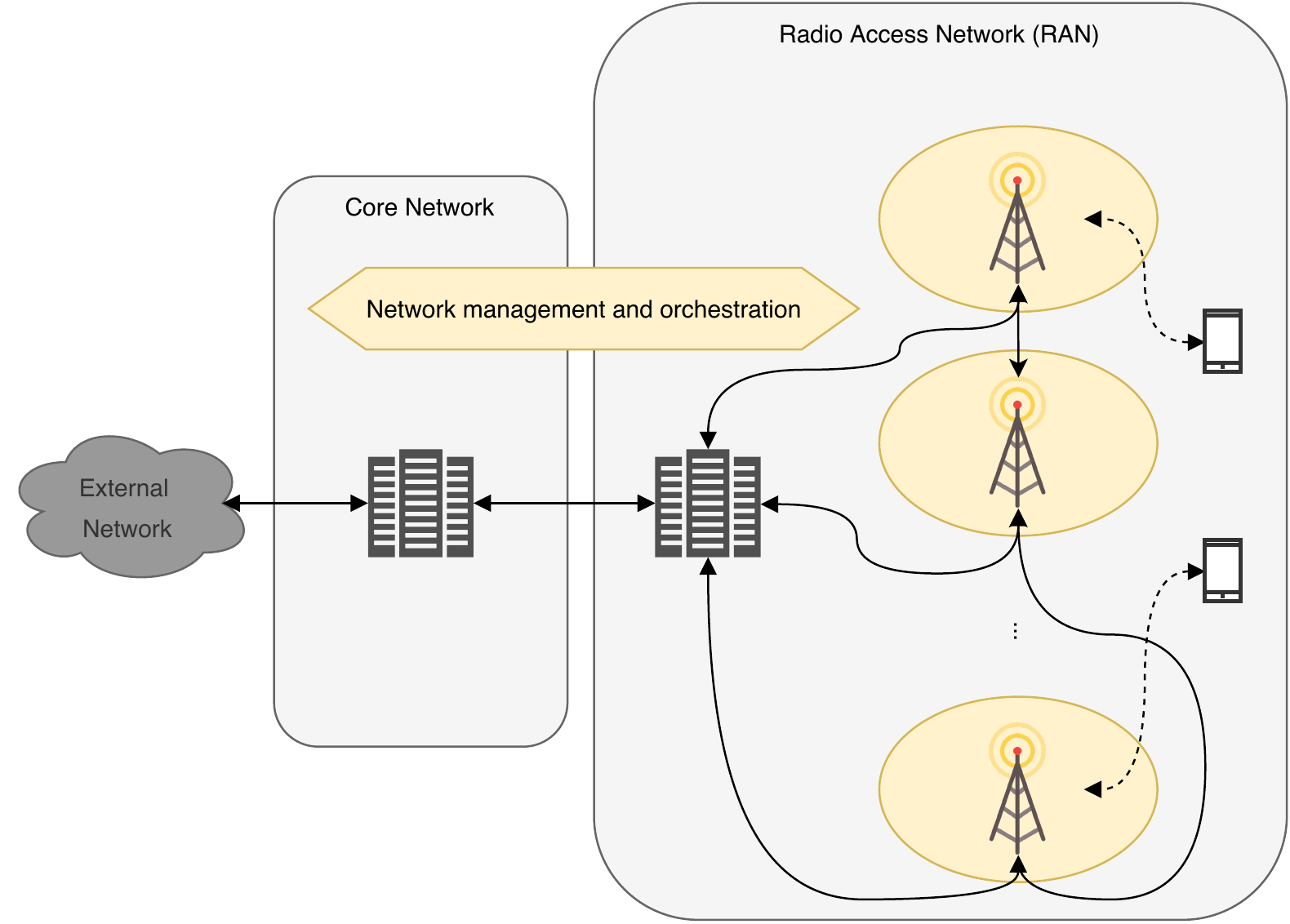}
	\caption{Mobile Network}
	\label{fig:mobile_network}
\end{figure}

A 5G mobile network (whether public or private) typically consists of a transport network (e.g., fiber optic network), a core network with central elements for network control, and the radio access network (RAN) that provides connections to mobile terminals. A schematic example of such a 5G mobile radio network can be seen in Figure \ref{fig:mobile_network}. While there is a plethora of vendors for virtualized core networks, radio access networks are provided by only a handful of major network equipment vendors. Today's RANs are in addition highly integrated solutions from individual vendors, with little interoperability between products from different vendors. This inevitably leads to innovation barriers.

A key to more innovation in mobile networks lies in the Open RAN approach, which uses RAN technologies based on disaggregation and openness. In the Open RAN approach, the RAN is divided into several RAN units, each of which performs different RAN functions. The crucial point here is that the interfaces between the RAN units are open and guarantee interoperability. The open interfaces are therefore the basis for more flexibility and the much needed trust in communication technologies. Finally, Open RAN promises performance enhancements over the current integrated vendor-specific solutions. 

 In addition to the architectural disaggregation and openness (in the sense of interoperability), the aspects of cloudification and virtualisation \cite{9124820}, network slicing \cite{9367527}, \cite{10.1109/MCOM.2018.1701319} and machine learning \cite{9647671} also play an important role in the Open RAN context. Yet, it is important to emphasize that except for disaggregation and openness, the other aspects such as virtualization and machine learning are not an inherent part of the Open RAN concept.  This means that Open RAN systems basically do not need to be virtualized, which can be beneficial in some cases.
 For instance, in Massive MIMO systems, it may be beneficial in terms of energy-efficiency to implement the lower layer RAN functions on a dedicated system-on-chip (SoC) rather than running these functions on general-purpose processors. Therefore, Open RAN is not equivalent to virtualized RAN (vRAN), even though many Open RAN systems are highly virtualized systems. However, this is also true for highly integrated and closed RAN systems.  
 
 For a better understanding of this paper, it is also helpful to distinguish between Open RAN, O-RAN, and OpenRAN (one word), as these terms are often confused or used interchangeably. The acronym O-RAN originates from the O-RAN Alliance, which focuses largely on the development of the O-RAN architecture. This architecture forms the basis for the analysis in this paper. OpenRAN (one word), on the other hand, is a project group established by the Telecom Infra Project (TIP). This term plays no role in this paper. Finally, Open RAN is used as a generic term for disaggregated systems with open and interoperable interfaces. The O-RAN architecture is one possible Open RAN architecture. Especially in in the context campus networks, there might be different flavors of the Open RAN architecture. 
 
 %The architecture's openness and the advent of new IT technologies in the radio access network (RAN) are highly promising when it comes towards fulfilling the requirements for our future RAN’s.
 
 %The main architectural requirements for OpenRAN relate to the uncoupling of hardware and software, cloud infrastructure, and standardised and open interfaces among network functions.

\subsection{Architectural Overview}
\label{architectural_overview}

\begin{figure}
	\centering
	\includegraphics[width=0.5\textwidth]{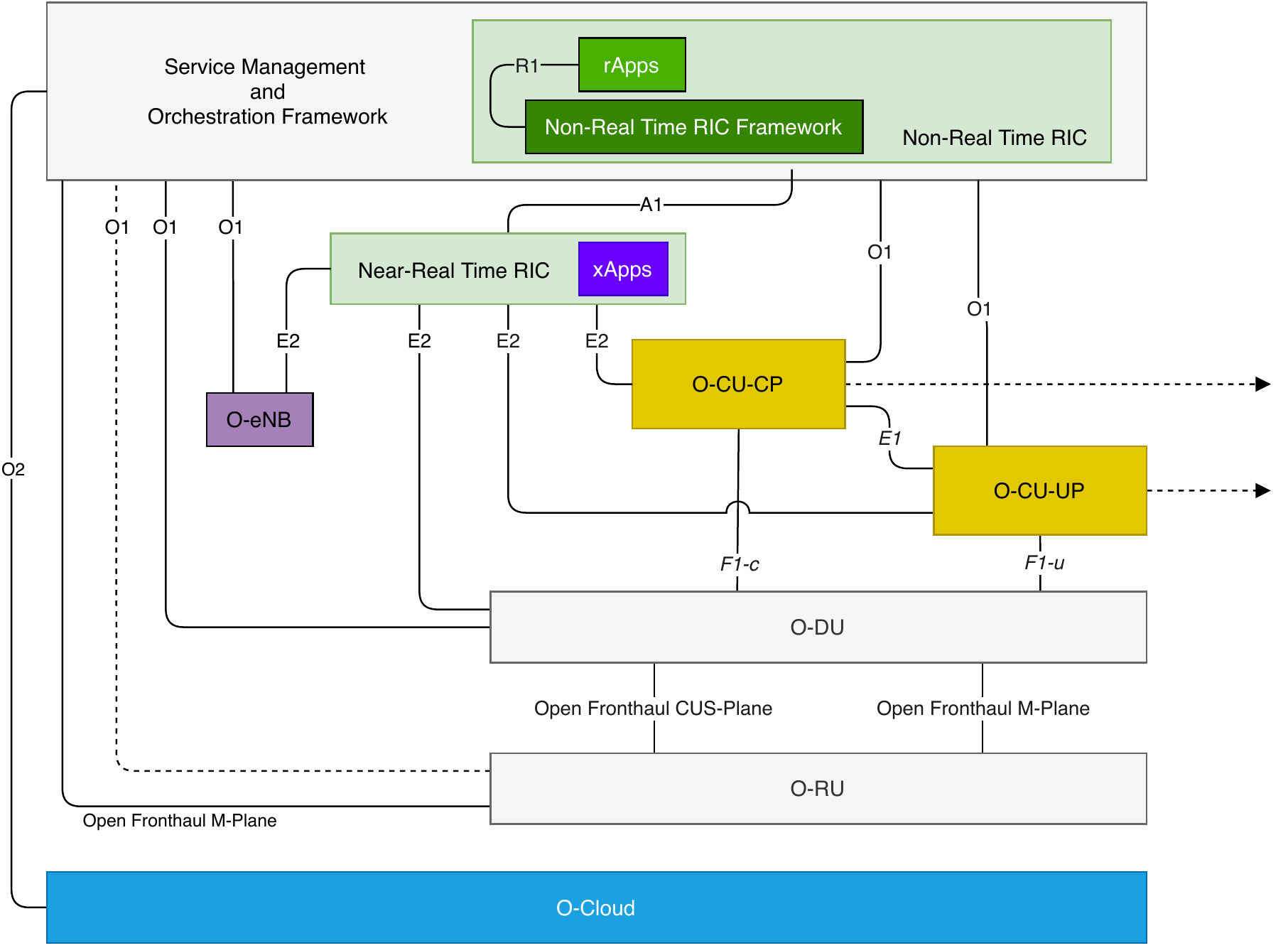}
	\caption{O-RAN specific interfaces \cite{wg1_workgroup_2021}}
	\label{fig:o_ran_interfaces}
\end{figure}

A traditional 3GPP-specified NG-RAN (Next Generation NodesBs Radio Access Network) is divided into two logical RAN units: The CU (Central Unit) and the DU (Distributed Unit). These basic units in turn comprise several logical functional units. All of them together are then connected to the core network. In the Open-RAN, the above two RAN functions are further divided according to the 3GPP definition. Figure \ref{fig:o_ran_interfaces} shows a simplified version outlining the breakdown of RAN functions and interfaces for a 3GPP compliant O-RAN architecture \cite{wg1_workgroup_2021}. 
\\New RAN functions specifically defined in the context of O-RAN are:

\begin{enumerate}
    \item Service Management and Orchestration (SMO) Framework
    \item RAN Intelligent Controllers (RICs) in the variants non real time (Non-RT RIC) and near real time (Near-RT RIC)
    \item Remote Unit (O-RU)
    \item O-Cloud
\end{enumerate}

We briefly explain the newly introduced interfaces within O-RAN in Section \ref{important_interfaces}.

\subsection{Security Advantages}
\label{security_advantages}

While most critics of Open-RAN architectures often criticise the general security or describe it as poor, we rather see the great opportunity to effectively apply security solutions across the board. It is of extreme importance in all projects to include security assessment at an early stage. As we are still in the initial design phase for O-RAN, we can now develop and incorporate security concepts in a timely manner. The concept of security by design is based on the fact that the impact of iterating an architecture in its early stages towards a secure design is many times more effective and leads to more stable outcomes than conducting a security analysis in the later stages of a technology's lifecycle \cite{sec_by_design}. A risk analysis \cite{BSI_2021_openran} has already been commissioned by the German Federal Office for Information Security (BSI). We have used this analysis as the basis for our further research and built on it.
The BSI study is an important contribution because it highlights the risks associated with the O-RAN architecture and has triggered an important discussion about security in 5G networks.  In this paper, we would like to focus more on the opportunities of the Open RAN approach and raise other important issues such as post-quantum security.

\section{Important Interfaces}
\label{important_interfaces}
%\todo[inline]{Currently only described briefly and concisely. Too superficial?}
%\todo[inline]{JK: for non-experts it is definitely too superficial. But non-experts are not our target audience I guess}

In this section we briefly mention the most important interfaces within an O-RAN deployment and summarise them. The descriptions of the individual interfaces are taken from the the official O-RAN specifications.

\subsection{O-Cloud}
\label{o_cloud}

The O-Cloud is a cloud-based computing platform that includes a collection of physical infrastructure nodes and hosts the components of relevant O-RAN functions (e.g., Near-RT RIC, O-CU-CP, O-CU-UP and O-DU). Beyond that, it also provides support for software components as well as the corresponding management and orchestration functions. Generally speaking it can be seen as the central execution environment for O-RAN components \cite{o_ran_architecture_desc6}.

\subsection{O1-Interface}
\label{o1_interface}

This interface grants access to network capabilities to the service management and orchestration framework. Here, network management is implemented according to the FCAPS model \cite{o_ran_oam_spec6}. FCAPS follows to the ISO model for telecommunications network management, which defines and incorporates the fault, configuration, accounting, performance, and security management task areas \cite{itu_m3010}, \cite{hegering1999}. 

%\todo[inline]{JK: add reference for FCAPS?}

\subsection{O2-Interface}
\label{o2_interface}

This interface plays a central role in the O-RAN environment.  It is a tool for managing and orchestrating open management and services. Its objective is to guarantee secure communication between the SMO framework and the O-Cloud platform. Therefore, the O2 interface as such is also extremely empowering \cite{o_ran_cad_wg6}.

\subsection{A1-Interface}
\label{a1_interface}

The A1-Interface facilitates communication between the Non-Real-Time RIC and the Near-Real-Time RIC. This involves the transmission of data from internally and externally O-RAN sources to the SMO-Framework. As an example, the declarative A1 policy-based guidelines, which contain statements about goals and resources for UEs and cells, can be mentioned here. Other administrative information shared via this interface are used for ML models (training, updating, use of ML models). Various internal as well as external O-RAN data sources are made available as enrichment information. The availability and general use of these sources is not crucial for the fulfilment of a task, but only serves the purpose of general improvement \cite{o_ran_architecture_desc6}. 
%\todo[inline]{IB: A1-Interface is missing in Fig. 	\ref{fig:o_ran_interfaces}}

\subsection{R1-Interface}
\label{r1_interface}

To access RIC functions that do not run in real time, the so-called rApps utilise the R1 interface. Examples here include the provision of policy-based guidance and enrichment information which is obtained through the A1 interface. In addition, data analysis, AI/ML training and information retrieval for RAN optimisation or for usage by other rApps is another main component. As well as the recommendation of configurations which can be transmitted via the O1 interface \cite{o_ran_nonrtric_wg2}.

\subsection{E2-Interface}
\label{e2_interface}

In order to connect the near-RT RIC with the so-called E2 nodes, the E2-Interface is used. In general it shall support all protocol layers and interfaces that are defined in 3GPP radio access networks. The objective is to manage and improve the E2 nodes and the resources they consume. To make this feasible, the "RAN Function Network Interfaces (NI)" service can observe and, if necessary, modify the entire data traffic of the network interface of each individual direct node \cite{o_ran_ic_e2m_wg3}.

\subsection{Open Fronthaul M-Plane}
\label{open_fronthaul_m_plane}

The Open Fronthaul M-Plane interface allows the management of the open radio unit components. It is used for performance reporting or for initialising and configuring operating parameters. Particularly relevant from a risk perspective is the possibility of updating the software of the components with this interface \cite{o_ran_mplane_wg4}.

\subsection{Open Fronthaul CUS-Plane}
\label{open_fronthaul_cus_plane}

The user and control plane data of the Uu interface are transmitted via the Open Fronthaul CUS-Plane interface. Furthermore, this component is in charge of synchronizing the time between the Open Distributed Unit and the Open Radio Unit. \cite{o_ran_cusplane_wg4}.

\section{Stakeholders}
%\todo[inline]{Should we add a figure for better explaining the relation between the actors/stakeholders}\todo[inline]{JK: yes}

% \begin{figure}[h]
% 	\centering
% 	\includegraphics[height=0.35\textwidth]{Figures/actors_graph.pdf}
% 	\caption{Hierarchy of the individual actors}
% 	\label{fig:actors}
% \end{figure}

To be able to analyse risks within an architecture more precisely, it is important to determine all stakeholders within the system and to define their capabilities. In table \ref{tab:stakeholder_listing}, possible attackers are therefore briefly explained according to the models defined by \cite{mimran_OPENRAN}, \cite{BSI_2021_openran} and \cite{NIS_openran}. We have extracted what we consider to be the most important stakeholders and categorised them according to their level of access and overall safety-critical impact.

In total, we have assigned four (from L1 to L4) access levels. The higher the level, the greater the potential impact of the capabilities available to the stakeholder. The categorisation of the individual levels was declared by us on the basis of the respective possibilities and their effects. All stakeholders who use only a few or only individual areas of the RAN and can exploit it if fall under L1. We therefore assume a relatively \textit{low} common risk. At level two, the stakeholder can further influence the component and also control it if necessary, i.e. he has explicit knowledge of the respective device and either operates it himself or knows to what extent and in what environment it is operated. In terms of risk, L2 already causes considerably more damage than L1, which is why we declare the overall risk to be \textit{medium}. The next level, L3, describes all stakeholder entities that have control and access to a complete stand-alone RAN system. This class poses a very high risk, as the instances have the ability to manipulate any services or hardware components. Thus, L3 creates a \textit{high} risk with regard to the entire system. The last and highest access category is level 4, representing the highest possible general risk (\textit{severe}). It refers to actors which are specialised in the construction or provision of RAN subsystems as well as complete systems. The main difference to L3 is that L4 stakeholders not only operate the network, but also configure and create it themselves. This means that they have fundamental knowledge about the functionality of the individual components. Thus, it can be assumed that possibly malicious L4 stakeholders can cause the greatest possible damage. Another point that underlines the seriousness of this is that L4 entities can already manipulate the overall RAN at an early stage. This means, for example, before the RAN operator in L3 has the chance to do so. It may be the case that an L3 stakeholder also falls into the L4 category at the same time, e.g. if an MNO integrates its RAN system completely itself. 

% \todo[inline]{IB: A short explanation of levels L1 to L4 would be helpful for a reader who is not an expert in system level security, i.e. me :) -> FK: You're right, I added a description of the individual levels.}
%\todo[inline]{IB: Again, it would be helpful to indicate/explain how the levels L1 to L4 are matched to the overall risk in the last column of Table \ref{tab:stakeholder_listing}. Maybe an explanation on an example of an actor like Government Services. -> FK: Makes sense to add one example, will finish that tomorrow. }

%To illustrate the relationship between the individual levels and the overall risk, we describe the relationship using the example of the government service provider. In this case, we assume that the state institution has the ability to tap legitimate network traffic and decode it into plaintext. 

Later, in connection with the interfaces explained in section \ref{important_interfaces}, we will derive which measures are necessary to reduce or mitigate the attack surfaces.
% \todo[inline]{IB: We should consolidate the usage of the words actor and stakeholder. In the table we use the word stakeholder while in the text of this section we use interchangeably actor and stakeholder. -> FK: Good point = DONE!}
\begin{table*}[h]
\caption{Itemised categorisation of the key stakeholder}
\label{tab:stakeholder_listing}
\renewcommand{\arraystretch}{1.2}
\begin{tabular}{|l|c|l|l|c|}
\hline
\textbf{Stakeholder}                                                                                              & \multicolumn{1}{l|}{\textbf{Access}} & \textbf{Short Description}                                                                                                                                                                                                                & \textbf{Capability}                                                                                                                                                                              & \multicolumn{1}{l|}{\textbf{Overall Risk}} \\ \hline
External \cite{NIS_openran} / Outsider \cite{BSI_2021_openran}                                                                                 & L1                                   & \begin{tabular}[c]{@{}l@{}}Does not match with the other \\ defined stakeholder classes. \\ Is able to access the interfaces \\ defined by 3GPP or the O-RAN \\ Alliance.\end{tabular}                                                    & \begin{tabular}[c]{@{}l@{}}Is capable of performing designated types \\ of threats in several risk zones\end{tabular}                                                                                  & \cellcolor[HTML]{FFCE93}Low                    \\ \hline
\begin{tabular}[c]{@{}l@{}}Network consumer\\ (Similar to User \cite{BSI_2021_openran} \\ \& Connected Devices \cite{NIS_openran})\end{tabular}     & L1                                   & \begin{tabular}[c]{@{}l@{}}Is participating in the network as \\ a normal and legitimate service \\ consumer by connecting \\ entities or services that facilitate \\ network functionality.\end{tabular}                   & \begin{tabular}[c]{@{}l@{}}Is able to exploit specific types of \\ threats in different risk zones using \\ legitimate credentials/secrets to exploit \\ the respective network.\end{tabular}    & \cellcolor[HTML]{FFCE93}Low                    \\ \hline
Government Services \cite{NIS_openran}                                                                                    & L2                                   & \begin{tabular}[c]{@{}l@{}}Authority that has legal rights \\ to intercept/tap network traffic. \end{tabular}                                                                                                                               & \begin{tabular}[c]{@{}l@{}}Decoding and reading of data sets sent \\ over the network\end{tabular}                                                                                               & \cellcolor[HTML]{FE996B}Medium                 \\ \hline
\begin{tabular}[c]{@{}l@{}}Hard-/Software Suppliers \cite{NIS_openran} \\ \& Manufacturers \cite{mimran_OPENRAN}\end{tabular} & L2                                   & \begin{tabular}[c]{@{}l@{}}Vendor of one or more specific \\ hard- or software component utilised in \\ the system "... providing services or\\ infrastructure to MNOs in order \\ to build and/or operate their networks" \cite{NIS_openran} \end{tabular} & \begin{tabular}[c]{@{}l@{}}Can substitute benign hard-/software \\ components with malicious ones\end{tabular}                                                                                      & \cellcolor[HTML]{FE996B}Medium                 \\ \hline
\begin{tabular}[c]{@{}l@{}}RAN-Operator \cite{BSI_2021_openran} \\ \& MNOs \cite{NIS_openran}\end{tabular}                                & L3                                   & \begin{tabular}[c]{@{}l@{}} Refers to traditional Mobile (Virtual) \\ Network operators, as well as critical \\ infrastructure operators from non \\ telecommunication sectors. \\ Has complete and comprehensive \\ control over the respective RAN. \\\end{tabular} & \begin{tabular}[c]{@{}l@{}}Possesses enhanced capabilities to \\ tamper hardware components as well as\\ specific services\end{tabular}                                                          & \cellcolor[HTML]{F8A102}High                   \\ \hline
RAN-Integrator                                                                                                    & L4                                   & \begin{tabular}[c]{@{}l@{}}Is specialised in assembling RAN \\ subsystems into a functioning entity \\ and ensuring that the subsystems \\ function smoothly together.\end{tabular}                                                       & \begin{tabular}[c]{@{}l@{}}Controls and operates software deployed \\ in the RAN as well as all hardware. \\ Thus, he has the possibility to \\ manipulate all individual entities at \\ each available stage.\end{tabular} & \cellcolor[HTML]{F56B00}Severe                 \\ \hline
\end{tabular}
\end{table*}

\section{Mitigations of Security Problems}
%\todo[inline]{Section-title rather inappropriate TODO: Define the headings of the Security Suggestions groups // Talk about}

In this section, we deal with the issue of how to solve the problems denounced and, above all, what steps are necessary to do so and how much effort is required. For this purpose, we identify security aspects that should be applied. 
%These also partly reflect the recommendations of the points mentioned in the BSI report.

\subsection{Enforcement of clear safety \& security concepts}
\label{enforcement_safety_concepts}
One of the main components to fully secure a system is the use of standardised mechanisms to ensure both safety and security. It would be necessary, for example, to create a clear definition of a clear rights and role concept with regard to the communication of interfaces and services. Furthermore, for separation concepts, firewall-friendly designs, minimisation of the effects of denial of service and the implementation of a zero trust model. To go one step further in defining possible measures, we have extracted the individual proposals from the enisa report \cite{enisa_2021_nfv} for network function virtualisation. These measures can also be applied holistically to an Open-RAN system. In Table 2 we have listed the best practices divided according to their level of applicability. If we look at the individual practices, it becomes clear how extensive our current security and safety options are in order to facilitate a secure Open-RAN deployment.

\subsection{Mandatory Encryption}
\label{mandatory_encryption}
Parts of the O-RAN definition currently require no or only weak encryption, which weakens the overall security of a potential Open-RAN deployment. For example, the encryption at the transport layer is only optional and legacy protocols that rely on the weak encryption are not forbidden by default. Consequently, it is recommended to increase the security of O-RAN by mandating strong encryption and not allowing old protocols. This would indeed increase the security of O-RAN. It is noteworthy however, that previous radio access networks have the same shortcomings as Open-RAN when it comes to encryption, and have suffered from security vulnerabilities as a result. %TODO:Cite
The partial lack of strong cryptography is thus not a threat introduced by O-RAN itself, but an ever existing threat to telecommunication networks, that can now be fixed in the O-RAN specification. Following the described \textit{security by design} approach, we strongly encourage strong cryptography should be mandatory. This would result in a major security improvement compared to the previous standards. 
%however, even with optional encryption uld even not doing so does introduce new security vulnerabilities, but rather would result in inheriting old ones.

\subsection{Post-Quantum Security}
\label{PQS}

Due to the dynamic development of quantum computers and their expected future ability to break currently used classical public-key schemes, using post-quantum schemes, i.e., quantum-resistant cryptographic schemes, should be at least recommended within O-RAN. 
Thus, while encryption at large should be mandatory, ”shall support” for post-quantum cryptography would be sufficient for now. 
We want to stress that classically encrypted data is also at risk before powerful quantum computers exist, since the encrypted data can be stored now and be decrypted once powerful quantum computers exist.
Hence, the integrators have to carefully analyze which data need only short-term protection and which data need to be protected for several decades, which would then require post-quantum protection.
Since the NIST process for standardizing post-quantum cryptographic schemes is on the verge of announcing first schemes that will be standardized, standardized PQC schemes will exist once 6G is being deployed.

We want to emphasize two points that should be paid attention to when developing post-quantum secure telecommunication protocols: 
First, O-RAN relies mainly on symmetric cryptography to ensure authentication and confidentiality of the data. So far, post-quantum security considerations for telecommunication are often based on the assumption that quantum computers mainly pose a serious threat to asymmetric cryptography \cite{mitchell2020impact,yang2020overview5g} - and thus, telecommunication protocols can ensure post-quantum security by doubling the key-size. Arguably though, this claim is not backed by proofs but rather by a truism that guides post-quantum security considerations; so in addition to replacing the existing asymmetric cryptography, O-RAN's symmetric cryptosystems need to be re-evaluated w.r.t. quantum resilience (proofs) as well to back up the claim that doubling the key size is indeed sufficient.
Second, the resource-constrained environment of telecommunication infrastructure (e.g., SIMcards have low resources available, low network bandwidth requirements are a must) and highly adversarial threat environment (e.g., SIM cards are highly susceptible to side channels, the need to defend against nation-station actors, and the fact that cloud environments introduce new threats) require to carefully evaluate and tailor post-quantum cryptography schemes towards use in telecommunication protocols. This process is time-intensive - given the slow nature of telecommunication standardization bodies and even more so the slow nature of actually implementing a new telecommunication standard, it is urgent to address the topic (and integrate post-quantum security in the standard) already now.  
Recent works have already started to explore these trade-offs for the Subscriber Concealed Identifier (SUCI), which aims at concealing subscriber identities \cite{ulitzschpqcsuci}. The developed post-quantum secure protocols for the SUCI, tailored towards use in telecommunication protocols, serve as a great starting point to explore the challenges and necessary changes required when bringing post-quantum security to next generation telecommunication networks.

\subsection{Cloud Environments}
\label{cloud_environments}
Moving Open-RAN components to the cloud results in a new threat landscape, specifically when considering a malicious cloud provider. Recent risk analysis correctly states that a cloud provider that controls the O-Cloud has the same capabilities as the RAN-Operator. Currently, there are few mandatory safety measures in the O-RAN requirements and therefore two recommendations can be clearly made to mitigate the problems: 1) Integrate security measures to defend against malicious cloud provider, for example, through Trusted Execution Environments and 2) to integrate mandatory access control and security requirements in the O-RAN definition.

A malicious cloud provider would indeed undermine a RAN's security. In practice however, operators expect to build and run their own data-centers instead of relying on external cloud solutions. As a result, the O-Cloud operator can be assigned the same level of trust as the RAN-Operator themselves. This completely mitigates the malicious cloud provider scenario.

We also recommend to only use trusted datacenters and cloud solution for the O-Cloud: Defending against malicious cloud providers through the usage of confidential computing and trusted execution environments is non-trivial: The security provided by confidential computing and trusted execution environments have been undermined with various attack vectors, stemming from the very powerful attacker model. 

If the O-Cloud is trusted and adheres to the standard security best practices in its configuration and design, we expect the security risk induced through a cloud-based RAN to be minimal.

\subsection{Clarification \& concrete definition}
\label{clarification_concrete_definition}
From the point of view of security research, the current state of the O-RAN specification still leaves a number of wide gaps in the specification of security aspects. One of the reasons why this is the case is the philosophy of the O-RAN Alliance to mainly provide some kind of guidelines. This is why the term "shall support" is often found in the documents. The actual implementation of the necessary security concepts is, in their view, the responsibility of the integrator or hardware manufacturer. Of course, this is a clear thorn in the side for an authority like the German Federal Office for Information Security, the BSI. The German Technical Inspection Agency (TÜV) would also like to have more concrete definitions, e.g., in order to be able to precisely evaluate and approve an Open-RAN system. However, this will probably never be possible with the plan pursued by the O-RAN Alliance, since they only provide a possible architecture and specifications for it, but do not develop an additional standard to 3GPP. The O-RAN Alliance merely defines a technical concept that is intended to improve interoperability in the radio access networks of mobile networks. This fact, however, offers us a great opportunity as security researchers in this field. We are now able to put an additional security view on top of the concept and thus mitigate all security-critical concerns. 
%\todo[inline]{Here we should perhaps consider what we as 6G-RIC want to officially contribute to this and also "announce" it here.}

\subsection{Privacy}
\label{Privacy}
Technological innovations and emerging trends in the context of 6G, such as the tighter integration/convergence of sensing and communication, may pose unique challenges to both communication security and privacy. For example, in dual radar and communication systems, the inclusion of data into the probing signal, used to illuminate targets, makes it prone to eavesdropping from potentially malicious targets. Even if the data itself is protected with higher-layer encryption, the existence of a communication link can still be detected from a malicious agent, thus making it prone to cyberattacks \cite{ISAC_Security}. Similarly, the introduction of sensing capabilities and explicit localization of users/devices to improve communication network performance, may pose significant privacy challenges.  

In face of these challenges, it is important to assess the potential of certain approaches to provide secure, privacy-preserving solutions at the radio access level. E.g. from a privacy perspective, it is essential to collect only user related data that is absolutely necessary for the operation of the network and to move away from the explicit localization paradigm as much as possible. In addition, appropriate privacy-enhancing technologies should be integrated, e.g. differential privacy.
An example is provided by the use of \textit{channel charting} \cite{CC_Studer} to enhance network functionalities such as, e.g., radio resource management, beam management (mmWave and sub-THz), cell-association and handover.  Channel charting uses unsupervised/semi-supervised learning to embed high-dimensional information about the radio environment into a low-dimensional chart and relies on a pseudo-location (i.e. location in the low-dimensional chart) of the users. The use of pseudo-location on a channel chart can be seen as a privacy protection feature, allowing localization-related services to be delivered without requiring the actual user location to be estimated. While promising in general, the concept needs to be further formalized and thoroughly investigated from the perspective of privacy before inclusion as a design metric. 

\section{Machine Learning}
\label{Machine Learning}

%\todo[inline]{Rough overview of machine learning mitigation in O-RAN. Here we lay the groundwork and give our opinion for a possible follow-up paper.}

O-RAN seeks to utilize novel Machine Learning (ML) techniques such as deep learning to automate operational network functions and reduce operational cost once applied at both component and network levels \cite{alliance2018ran}. However, the Open-RAN architecture entails security challenges because of its inherently open and modular nature. Devices within Open-RAN are able to run software that does not trust the hardware it is running on. Due to the openness of Open-RAN, it is susceptible to intrusion. Cyber-attacks pose a threat to security goals and could result in denial-of-service (DoS) within the Open-RAN network. These attacks present vulnerability to the Non-RT RIC and the Near-RT RIC controller operations within Open-RAN \cite{masur2021artificial}.
% possible subtitle: anomaly detection
\subsection{Anomaly Detection}
Ideally, intrusions are detected with high performance, high speed, and a low false-positive alarm rate. It is therefore clear that traditional human means will not be enough to detect and combat cyber-attacks in the required manner. Consequently, robust ML techniques will also play a big role in tackling anomaly-based intrusion detection. \newline
Anomaly-based IDS can be divided into three classes: statistical anomaly IDS, knowledge-based IDS, and ML IDS \cite{kocher2021machine}. The focus here is on novel ML-based IDS. The ML-based anomaly detection methods in IDS \cite{wang2021machine} can be classified into supervised learning, semi-supervised learning, unsupervised learning, reinforcement learning, and graph neural networks \cite{ma2021comprehensive}. The limitations of traditional (shallow) ML-based IDS, such as the reliance on manual feature engineering to extract useful information from network traffic and dealing with unlabeled, high-dimensional data, have paved the way for Deep Learning (DL)-based IDS \cite{liu2019machine} that do not require manual feature engineering and can automatically learn complex features from raw data due to their deeper structure \cite{ahmad2021network}.\newline 
Initial RAN-specific anomaly detection approaches have been presented by \cite{momkute2018adapted}, \cite{yuan2020anomaly}, \cite{sundqvist2022uncovering}, \cite{mismar2021unsupervised}. However, the O-RAN architecture with its different interfaces poses new challenges for an anomaly detector, as next-generation RAN data can be roughly divided into performance management, configuration management, and fault management data. It has been shown that considering only performance management data in anomaly detection can lead to sub-optimal results \cite{sundqvist2022uncovering}, \cite{mismar2021unsupervised}. Therefore, it is important to define the requirements under which the anomaly detector must operate, as not all approaches presented so far take into account the different data streams in the next-generation RAN. \newline In terms of where the anomaly detector is deployed, for example, the training of ML models can be performed in the Non-RT RIC controller. Subsequently, the learned model is inputted into the Near-RT RIC controller which uses this model on Real-Time data and makes Real-Time decisions in an online manner \cite{ali20206g}.\newline
% possible subtitle: importance of robustness

\subsection{Importance of Robustness}
Moreover, every ML-based learning method mentioned above is prone to an attack technique referred to as adversarial ML \cite{LiuaAdversarialWireless}. Hence, incorporating ML solutions into the RAN poses new cyber-security threats. Consequently, having a thorough understanding of the attack surface and ensuring the robustness of the O-RAN to adversarial machine learning threats are mandatory for securing the new Open-RAN architecture \cite{bitton2022adversarial}.\newline

% possible subtitle: Explainable AI
\subsection{Explainable AI}
Notably, Explainable AI (XAI) has a great potential for securing ML systems as explanations are key in identifying and defending against different types of attacks. To elaborate, if explanations become available for adversarial attacks, they become easier to defend against \cite{rawal2021recent}. Additionally, explanations can support effective root cause analysis and localisation \cite{chawla2020interpretable}.

\begin{table*}
\label{table_practices}
\centering
\caption{List of best practices presented by enisa \cite{enisa_2021_nfv}}
\begin{tabular}{|c|l|c|l|} 
\hline
Level                           & \multicolumn{1}{c|}{Practice}                              & Level                             & \multicolumn{1}{c|}{Practice}                                  \\ 
\hline
\multirow{2}{*}{Organisational} & Trust model                                                & \multirow{26}{*}{Technical ~ ~ ~} & Tracking version changes                                       \\ 
\cline{2-2}\cline{4-4}
                                & SLAs establishment                                         &                                   & Deployment security                                            \\ 
\cline{1-2}\cline{4-4}
\multirow{16}{*}{Policy}        & Zero Trust                                                 &                                   & Software detection or relocation                               \\ 
\cline{2-2}\cline{4-4}
                                & Security Assessment of new or changes to existing Services &                                   & (Post-Quantum) Cryptography                                                   \\ 
\cline{2-2}\cline{4-4}
                                & Vulnerability handling  patch management                   &                                   & Hypervisor protection                                          \\ 
\cline{2-2}\cline{4-4}
                                & Security testing and assurance                             &                                   & Security Management and orchestration                          \\ 
\cline{2-2}\cline{4-4}
                                & Incident management                                        &                                   & Remote attestation                                             \\ 
\cline{2-2}\cline{4-4}
                                & Secure Update Management                                   &                                   & Software compliance and integrity preservation                 \\ 
\cline{2-2}\cline{4-4}
                                & Restriction on installing applications                     &                                   & Segmentation/isolation between network functions  \\ 
\cline{2-2}\cline{4-4}
                                & Defense in depth                                           &                                   & Secure boot integrity                                          \\ 
\cline{2-2}\cline{4-4}
                                & String password policy                                     &                                   & Data protection and privacy                                    \\ 
\cline{2-2}\cline{4-4}
                                & Secure supply chain                                        &                                   & Encrypting Volume/swap Areas                                   \\ 
\cline{2-2}\cline{4-4}
                                & Resources inventory management system and database         &                                   & Trusted computing technologies                                 \\ 
\cline{2-2}\cline{4-4}
                                & Apply hardening policies                                   &                                   & Hardware security                                              \\ 
\cline{2-2}\cline{4-4}
                                & Multi-vendors segregation and trust                        &                                   & Centralised log auditing                                       \\ 
\cline{2-2}\cline{4-4}
                                & Security by design                                         &                                   & Use and ownership of "root" administration credentials         \\ 
\cline{2-2}\cline{4-4}
                                & Life cycle management                                      &                                   & Local or removal Blade Storage - SAN protection                \\ 
\cline{2-2}\cline{4-4}
                                & Software Bill Of Materials (SBM)                           &                                   & Network security                                               \\ 
\cline{1-2}\cline{4-4}
\multirow{8}{*}{Technical}      & Trusted time source                                        &                                   & SDN security management                                        \\ 
\cline{2-2}\cline{4-4}
                                & Secure 3rd party hosting environments                      &                                   & MANO access control and management                             \\ 
\cline{2-2}\cline{4-4}
                                & Redundancy and backup                                      &                                   & VIM/CISM connectivity to Hypervisor/CIS                        \\ 
\cline{2-2}\cline{4-4}
                                & Specific container security controls                       &                                   & Recovery and reinstallation                                    \\ 
\cline{2-2}\cline{4-4}
                                & OSS/BSS protection                                         &                                   & Deploying VMs/Containers of differing trust levels             \\ 
\cline{2-2}\cline{4-4}
                                & LI capabilities                                            &                                   & Orchestration platform security management                     \\ 
\cline{2-2}\cline{4-4}
                                & User plane security                                        &                                   &                                                                \\ 
\cline{2-2}
                                & MEC security                                               &                                   &                                                                \\
\hline
\end{tabular}
\end{table*}

\section{Conclusion}
In our research so far, we have not found that Open-RAN concepts such as O-RAN introduce major problematic security issues. Of course, there is an increased attack target due to the larger surface area of the ecosystem. However, such complex systems can be secured through various practices such as in Table 2. We can only agree with the opinion of Mimran et. al. \cite{eval_sec_o_ran} that the security risks that arise are mainly due to the 5G requirements and less due to the specific decisions in the Open-RAN architecture. In general, we can say that now with O-RAN the attack surface is much clearer than in contrast to previous proprietary implementations, which were unclear and in general rather a big question mark. Our task now is to define the overall security methodologies for the individual critical points and to apply them. This will then provide integrators and network operators with options for operating their Open-RAN deployment securely.

\section*{Acknowledgment}
 The authors acknowledge the financial support by the Federal Ministry of Education and Research of Germany in the programme of ``Souverän. Digital. Vernetzt.'' Joint project 6G-RIC, project identification number: 16KISK020K and 16KISK0[21-35].
 
 The authors of this paper also acknowledge the support of educational conversations with Jean-Pierre Seifert around the topics discussed in this paper, as well as the German Federal Office for Information Security (\textit{BSI}).

\printbibliography[heading=bibintoc]

@inproceedings{ulitzschpqcsuci,
  title={A Post-Quantum Secure Subscription Concealed Identifier for 6G},
  author={Ulitzsch, Vincent and Park, Shinjo and Marzougui, Soundes and Seifert, Jean-Pierre},
  booktitle={To appear in 15th ACM Conference on Security and Privacy in Wireless and Mobile Networks (WiSec 2022)},
  year={2022}
}

@article{mitchell2020impact,
	title={The impact of quantum computing on real-world security: A 5G case study},
	author={Mitchell, Chris J},
	journal={Computers \& Security},
	volume={93},
	pages={101825},
	year={2020},
	publisher={Elsevier}
}

@article{yang2020overview5g,
	title={An overview of cryptographic primitives for possible use in 5G and beyond},
	author={Yang, Jing and Johansson, Thomas},
	journal={Science China Information Sciences},
	volume={63},
	number={12},
	pages={1--22},
	year={2020},
	publisher={Springer}
}

@INPROCEEDINGS{9367527,
  author={Kukliński, Sławomir and Tomaszewski, Lechosław and Kołakowski, Robert},
  booktitle={2020 IEEE Globecom Workshops (GC Wkshps}, 
  title={On O-RAN, MEC, SON and Network Slicing integration}, 
  year={2020},
  volume={},
  number={},
  pages={1-6},
  doi={10.1109/GCWkshps50303.2020.9367527}}

@article{10.1109/MCOM.2018.1701319,
author = {Elayoubi, Salah Eddine and Jemaa, Sana Ben and Altman, Zwi and Galindo-Serrano, Ana},
title = {5G RAN Slicing for Verticals: Enablers and Challenges},
year = {2019},
issue_date = {January 2019},
publisher = {IEEE Press},
volume = {57},
number = {1},
issn = {0163-6804},
url = {https://doi.org/10.1109/MCOM.2018.1701319},
doi = {10.1109/MCOM.2018.1701319},
journal = {Comm. Mag.},
month = {jan},
pages = {28–34},
numpages = {7}
}

@INPROCEEDINGS{9647671,  author={Vardakas, John S. and Ramantas, Kostas and Datsika, Eftychia and Payaró, Miquel and Pollin, Sofie and Vinogradov, Evgenii and Varvarigos, Manos and Kokkinos, Panagiotis and González-Sánchez, Roberto and Olmos, Juan Jose Vegas and Chochliouros, Ioannis and Chanclou, Philippe and Samarati, Pierangela and Flizikowski, Adam and Rahman, Md Arifur and Verikoukis, Christos},  booktitle={2021 IEEE International Mediterranean Conference on Communications and Networking (MeditCom)},   title={Towards Machine-Learning-Based 5G and Beyond Intelligent Networks: The MARSAL Project Vision},   year={2021},  volume={},  number={},  pages={488-493},  doi={10.1109/MeditCom49071.2021.9647671}}

@INPROCEEDINGS{9124820,  author={Singh, Sameer Kumar and Singh, Rohit and Kumbhani, Brijesh},  booktitle={2020 IEEE Wireless Communications and Networking Conference Workshops (WCNCW)},   title={The Evolution of Radio Access Network Towards Open-RAN: Challenges and Opportunities},   year={2020},  volume={},  number={},  pages={1-6},  doi={10.1109/WCNCW48565.2020.9124820}}

@misc{wg1_workgroup_2021, title={O-RAN Architecture Description 5.0, Technical Specification}, url={https://www.o-ran.org/specifications}, journal={WG1: Use Cases and Overall Architecture Workgroup}, year={2021}, month={Jul}}

@misc{o_ran_architecture_desc6, title={O-RAN Architecture-Description 6.0, Technical Specification}, url={https://www.o-ran.org/specifications}, journal={WG1: Use Cases and Overall Architecture Workgroup}, year={2022}, month={Mar}}

@misc{o_ran_oam_spec6, title={O-RAN Operations and Maintenance Interface Specification 6.0, Technical Specification}, url={https://www.o-ran.org/specifications}, journal={WG10: OAM for O-RAN}, year={2022}, month={Mar}}

@misc{o_ran_cad_wg6, title={Cloud Architecture and Deployment Scenarios for O-RAN Virtualized RAN v02.02}, url={https://www.o-ran.org/specifications}, journal={WG6: Cloudification and Orchestration Workgroup}, year={2021}, month={Oct}}

@misc{o_ran_nonrtric_wg2, title={O-RAN Non-RT RIC Architecture 1.0}, url={https://www.o-ran.org/specifications}, journal={WG2: Non-real-time RAN Intelligent Controller and A1 Interface Workgroup}, year={2021}, month={Oct}}

@misc{o_ran_ic_e2m_wg3, title={Near-Real-time RAN Intelligent Controller Architecture and E2 General Aspects and Principles v02.01}, url={https://www.o-ran.org/specifications}, journal={WG3: Near-real-time RIC and E2 Interface Workgroup}, year={2022}, month={Mar}}

@misc{o_ran_mplane_wg4, title={Near-Real-time RAN Intelligent Management Plane Specification 8.0}, url={https://www.o-ran.org/specifications}, journal={WG4: Open Fronthaul Interfaces Workgroup}, year={2022}, month={Mar}}

@misc{o_ran_cusplane_wg4, title={Fronthaul Control, User and Synchronization Plane Specification 8.0}, url={https://www.o-ran.org/specifications}, journal={WG4: Open Fronthaul Interfaces Workgroup}, year={2022}, month={Mar}}

@article{sec_by_design,
author = {Benzel, Terry and Irvine, Cynthia and Levin, Timothy and Bhaskara, Ganesha and Nguyen, Thuy and Clark, Paul},
year = {2005},
month = {01},
pages = {},
title = {Design Principles for Security}
}

@article{eval_sec_o_ran,
  author    = {Dudu Mimran and
               Ron Bitton and
               Yehonatan Kfir and
               Eitan Klevansky and
               Oleg Brodt and
               Heiko Lehmann and
               Yuval Elovici and
               Asaf Shabtai},
  title     = {Evaluating the Security of Open Radio Access Networks},
  journal   = {CoRR},
  volume    = {abs/2201.06080},
  year      = {2022},
  url       = {https://arxiv.org/abs/2201.06080},
  eprinttype = {arXiv},
  eprint    = {2201.06080},
  bibsource = {dblp computer science bibliography, https://dblp.org}
}

@article{alliance2018ran,
  title={O-RAN: towards an open and smart RAN},
  author={Alliance, Open RAN},
  journal={White paper},
  pages={1--19},
  year={2018}
}

@article{masur2021artificial,
  title={Artificial Intelligence in Open Radio Access Network},
  author={Masur, Paul H and Reed, Jeffrey H},
  journal={arXiv preprint arXiv:2104.09445},
  year={2021}
}

@article{kocher2021machine,
  title={Machine learning and deep learning methods for intrusion detection systems: recent developments and challenges},
  author={Kocher, Geeta and Kumar, Gulshan},
  journal={Soft Computing},
  volume={25},
  number={15},
  pages={9731--9763},
  year={2021},
  publisher={Springer}
}

@article{wang2021machine,
  title={Machine Learning in Network Anomaly Detection: A Survey},
  author={Wang, Song and Balarezo, Juan Fernando and Kandeepan, Sithamparanathan and Al-Hourani, Akram and Chavez, Karina Gomez and Rubinstein, Benjamin},
  journal={IEEE Access},
  volume={9},
  pages={152379--152396},
  year={2021},
  publisher={IEEE}
}

@article{ma2021comprehensive,
  title={A comprehensive survey on graph anomaly detection with deep learning},
  author={Ma, Xiaoxiao and Wu, Jia and Xue, Shan and Yang, Jian and Zhou, Chuan and Sheng, Quan Z and Xiong, Hui and Akoglu, Leman},
  journal={IEEE Transactions on Knowledge and Data Engineering},
  year={2021},
  publisher={IEEE}
}

@article{liu2019machine,
  title={Machine learning and deep learning methods for intrusion detection systems: A survey},
  author={Liu, Hongyu and Lang, Bo},
  journal={applied sciences},
  volume={9},
  number={20},
  pages={4396},
  year={2019},
  publisher={Multidisciplinary Digital Publishing Institute}
}

@article{ahmad2021network,
  title={Network intrusion detection system: A systematic study of machine learning and deep learning approaches},
  author={Ahmad, Zeeshan and Shahid Khan, Adnan and Wai Shiang, Cheah and Abdullah, Johari and Ahmad, Farhan},
  journal={Transactions on Emerging Telecommunications Technologies},
  volume={32},
  number={1},
  pages={e4150},
  year={2021},
  publisher={Wiley Online Library}
}

@inproceedings{momkute2018adapted,
  title={Adapted Anomaly Detection for RAN Performance},
  author={Momkute, Dovile and {\v{Z}}vinys, Karolis and Barzd{\.e}nas, Vaidotas},
  booktitle={2018 IEEE 6th Workshop on Advances in Information, Electronic and Electrical Engineering (AIEEE)},
  pages={1--4},
  year={2018},
  organization={IEEE}
}

@inproceedings{yuan2020anomaly,
  title={Anomaly Detection and Root Cause Analysis Enabled by Artificial Intelligence},
  author={Yuan, Yannan and Yang, Jiaolong and Duan, Ran and Chih-Lin, I and Huang, Jinri},
  booktitle={2020 IEEE Globecom Workshops (GC Wkshps},
  pages={1--6},
  year={2020},
  organization={IEEE}
}

@inproceedings{sundqvist2022uncovering,
  title={Uncovering latency anomalies in 5G RAN-A combination learner approach},
  author={Sundqvist, Tobias and Bhuyan, Monowar and Elmroth, Erik},
  booktitle={2022 14th International Conference on COMmunication Systems \& NETworkS (COMSNETS)},
  pages={621--629},
  year={2022},
  organization={IEEE}
}

@article{mismar2021unsupervised,
  title={Unsupervised learning in next-generation networks: Real-time performance self-diagnosis},
  author={Mismar, Faris B and Hoydis, Jakob},
  journal={IEEE Communications Letters},
  volume={25},
  number={10},
  pages={3330--3334},
  year={2021},
  publisher={IEEE}
}

@article{ali20206g,
  title={6G white paper on machine learning in wireless communication networks},
  author={Ali, Samad and Saad, Walid and Rajatheva, Nandana and Chang, Kapseok and Steinbach, Daniel and Sliwa, Benjamin and Wietfeld, Christian and Mei, Kai and Shiri, Hamid and Zepernick, Hans-J{\"u}rgen and others},
  journal={arXiv preprint arXiv:2004.13875},
  year={2020}
}

@article{bitton2022adversarial,
  title={Adversarial Machine Learning Threat Analysis in Open Radio Access Networks},
  author={Bitton, Ron and Avraham, Dan and Klevansky, Eitan and Mimran, Dudu and Brodt, Oleg and Lehmann, Heiko and Elovici, Yuval and Shabtai, Asaf},
  journal={arXiv preprint arXiv:2201.06093},
  year={2022}
}

@article{rawal2021recent,
  title={Recent Advances in Trustworthy Explainable Artificial Intelligence: Status, Challenges and Perspectives},
  author={Rawal, Atul and Mccoy, James and Rawat, Danda B and Sadler, Brian and Amant, Robert},
  journal={IEEE Transactions on Artificial Intelligence},
  volume={1},
  number={01},
  pages={1--1},
  year={2021},
  publisher={IEEE Computer Society}
}

@inproceedings{chawla2020interpretable,
  title={Interpretable Unsupervised Anomaly Detection For RAN Cell Trace Analysis},
  author={Chawla, Ashima and Jacob, Paul and Feghhi, Saman and Rughwani, Devashish and van der Meer, Sven and Fallon, Sheila},
  booktitle={2020 16th International Conference on Network and Service Management (CNSM)},
  pages={1--5},
  year={2020},
  organization={IEEE}
}

@ARTICLE{LiuaAdversarialWireless,

  author={Liu, Jinxin and Nogueira, Michele and Fernandes, Johan and Kantarci, Burak},

  journal={IEEE Communications Surveys   Tutorials}, 

  title={Adversarial Machine Learning: A Multilayer Review of the State-of-the-Art and Challenges for Wireless and Mobile Systems}, 

  year={2022},

  volume={24},

  number={1},

  pages={123-159},

  doi={10.1109/COMST.2021.3136132}}

@book{hegering1999, author = {Hegering, Heinz-Gerd and Abeck, Sebastian and Neumair, Bernhard}, title = {Integrated Management of Networked Systems: Concepts, Architectures, and Their Operational Application}, year = {1999}, isbn = {1558605711}, publisher = {Morgan Kaufmann Publishers Inc.}, address = {San Francisco, CA, USA} }

@misc{itu_m3010,
	title = {ITU-T Recommendation M.3010 : {Principles} for a telecommunications management network},
	url = {https://www.itu.int/rec/T-REC-M.3010},
	urldate = {2022-03-28},
}

@misc{ISAC_Security,
  doi = {10.48550/ARXIV.2107.07735},
  
  url = {https://arxiv.org/abs/2107.07735},
  
  author = {Wei, Zhongxiang and Liu, Fan and Masouros, Christos and Su, Nanchi and Petropulu, Athina P.},
  
  title = {Towards Multi-Functional 6G Wireless Networks: Integrating Sensing, Communication and Security},
  
  publisher = {arXiv},
  
  year = {2021},
  
  copyright = {arXiv.org perpetual, non-exclusive license}
}

@ARTICLE{CC_Studer,

  author={Studer, Christoph and Medjkouh, SaïD and Gonultaş, Emre and Goldstein, Tom and Tirkkonen, Olav},

  journal={IEEE Access}, 

  title={Channel Charting: Locating Users Within the Radio Environment Using Channel State Information}, 

  year={2018},

  volume={6},

  number={},

  pages={47682-47698},

  doi={10.1109/ACCESS.2018.2866979}}

@misc{BSI_2021_openran,
  author = {BSI},
  month = {11},
  title = {Open-RAN Risikoanalyse},
  url = {https://www.bsi.bund.de/SharedDocs/Downloads/DE/BSI/Publikationen/Studien/5G/5GRAN-Risikoanalyse.html},
  urldate = {2022-03-29},
  year = {2021},
  organization = {Bundesamt für Sicherheit in der Informationstechnik}
}

@misc{enisa_2021_nfv,
  author = {enisa},
  month = {2},
  title = {NFV Security in 5G - Challenges and Best Practices},
  url = {https://www.enisa.europa.eu/publications/nfv-security-in-5g-challenges-and-best-practices},
  urldate = {2022-03-29},
  year = {2022},
  organization = {European Union Agency for Cybersecurity}
}

@misc{NIS_openran,
  author = {NIS},
  month = {5},
  title = {Report on the cybersecurity of Open RAN},
  url = {https://digital-strategy.ec.europa.eu/en/library/cybersecurity-open-radio-access-networks},
  urldate = {2022-06-01},
  year = {2022},
  organization = {NIS Cooperation Group}
}

@misc{mimran_OPENRAN,
  doi = {10.48550/ARXIV.2201.06080},
  
  url = {https://arxiv.org/abs/2201.06080},
  
  author = {Mimran, Dudu and Bitton, Ron and Kfir, Yehonatan and Klevansky, Eitan and Brodt, Oleg and Lehmann, Heiko and Elovici, Yuval and Shabtai, Asaf},
  
  title = {Evaluating the Security of Open Radio Access Networks},
  
  publisher = {arXiv},
  
  year = {2022},
  
  copyright = {Creative Commons Attribution 4.0 International}
}
\vspace{12pt}

\begin{table}[t]
    \centering
    \label{table:abbrevations}
    \caption{List of abbreviations}
    \begin{tabular}{|l|l|}
    \hline
    \multicolumn{1}{|c|}{\textbf{Term}} & \multicolumn{1}{c|}{\textbf{Description}}   \\ \hline
    3GPP                                & Third Generation Partnership Project        \\ \hline
    BSS                                 & Business Support System                     \\ \hline
    CISM                                & Container Infrastructure Service Management \\ \hline
    CU                                  & Central Unit                                \\ \hline
    DU                                  & Distributed Unit                            \\ \hline
    DL                                  & Deep Learning                           \\ \hline
    DoS                                 & Denial-of-Service                           \\ \hline
    FCAPS                               & Fault, Configuration, Accounting, Performance and Security                            \\ \hline
    IDS                                 & Intrusion Detection Systems                        \\ \hline
    ISO                                 & International Organization for Standardization                         \\ \hline
    LI                                  & Lawful Interception                         \\ \hline
    MANO                                & Management and Orchestration                \\ \hline
    MEC                                 & Multi-access Edge Computing                 \\ \hline
    MIMO                                & Multiple Input Multiple Output              \\ \hline
    MNO                                 & Mobile Network Operator                       \\ \hline
    NG-RAN                              &  Next Gerneration NodesBs Radio Access Network \\ \hline
    OSS                                 & Operations Support System                   \\ \hline
    PQS                                 & Post-Quantum Security                    \\ \hline
    RAN                                 & Radio Access Network                        \\ \hline
    RIC                                 & RAN Intelligent Controller                  \\ \hline
    RU                                  & Remote Unit                                 \\ \hline
    SAN                                 & Storage Area Network                        \\ \hline
    SDN                                 & Software Defined Network                    \\ \hline
    SMO                                 & Service Management and Orchestration        \\ \hline
    SLA                                 & Service-level Agreement                     \\ \hline
    SoC                                 & System-on-Chip                              \\ \hline
    UE                                  & User Equipment                              \\ \hline
    VIM                                 & Virtual Infrastructure Manager              \\ \hline
    VM                                  & Virtual Manager                             \\ \hline
    vRAN                                & Virtualized Radio Access Network            \\ \hline
    XAI                                 & Explainable AI            \\ \hline
    \end{tabular}
\end{table}

\end{document}